\documentstyle[aps,multicol,pre,psfig]{revtex}
\draft
\begin{document}
\title{Boundary Spatiotemporal Correlations in a Self--Organized
Critical Model of Punctuated Equilibrium}
\author{Emma Montevecchi}
\address{Laboratorium voor Vaste Stoffysica en Magnetisme,
Katholieke Universiteit Leuven,\\ Celesteijnenlaan 200D, B-3001 Leuven, Belgium}
\author{Attilio L. Stella}
\address{Abdus Salam ICTP, P.O. Box 586, 34100 Trieste, Italy\\
INFM-Dipartimento di Fisica e Sezione INFN, Universita' di Padova,
I-35131 Padova, Italy}
\maketitle

\begin{abstract}
In a semi--infinite geometry, a
$1D$, $M$--component model of biological evolution realizes microscopically
an inhomogeneous
branching process for $M \to \infty$. This
implies in particular a size distribution exponent $\tau'=7/4$
for avalanches starting at a free end of the evolutionary chain. A bulk--like
behavior with $\tau'=3/2$ is restored if ``conservative''
boundary conditions strictly fix to its critical, bulk value
the average number of species directly involved in an evolutionary avalanche
by the mutating species located at the chain end.
A two--site correlation function exponent ${\tau_R}'=4$ is also calculated
exactly in the ``dissipative'' case,
when one of the points is at the border. These results, together with
accurate numerical determinations of the time recurrence exponent
$\tau_{first}'$, show also that, no matter whether dissipation is present
or not, boundary avalanches have the same
space and time fractal dimensions as in the bulk, and their
distribution exponents obey the basic scaling laws holding there.

\end{abstract}

\begin{multicols}{2} \narrowtext

\section{Introduction.}
Nature offers many examples of systems driven by some external force
towards an out of equilibrium state characterized by critical spatiotemporal
correlations\cite{BTW}. In this stationary state the accumulated stress is
dissipated by avalanches of activity which occur intermittently and cover all
space and time scales. Models of nonequilibrium critical dynamics
displaying such features have been proposed for several phenomena,
ranging from earthquakes\cite{Olami} to interface depinning\cite{Sneppen},
or biological evolution in ecosystems\cite{BS}.

Some models of this self--organized criticality (SOC) are characterized by
extremal dynamics. Among these, the model of biological evolution
introduced by Bak and Sneppen (BS) constitutes
an important example \cite{BS}. Especially for system with extremal dynamics,
very few exact results are available so far\cite{Maslov}.
Most of our insight is based on numerical simulations, scaling arguments
\cite{PMB},
or mean field solutions\cite{MF}, related in general to random
neighbor versions of the models.

Among the existing mean field approaches, a particularly rich and complete
one, proposed recently, allows to describe in terms of an inhomogeneous
branching process (BP) several properties of avalanches, including some
due to border effects\cite{CTS}. In view of its phenomenological
character, an
open interesting problem within such an approach remains the identification
of specific microscopic models realizing the scalings of the inhomogeneous BP
in some appropriate random neighbor, or similar limit\cite{Lise}.

A step towards the establishment of an analytical theory
of scaling in extremal dynamics systems was made recently
by Boettcher and Paczuski\cite{BP}, who computed exactly a correlation
function
of an $M$--component version of the BS evolution model in the limit when
$M$ approaches infinity. This result,
combined with numerical ones, allowed verification of
important general scaling laws for self organized critical behavior\cite{PMB}.
Such laws describe the connection between space and time fractal properties
of avalanches in the bulk.

For models like sandpiles, SOC can be established
thanks to effects of the boundary dissipation, which balanches the flux of
added particles\cite{BTW}. This basic circumstance,
together with experience with standard criticality, called recently
attention on the boundary scaling properties of avalanches
\cite{STC,Priezzhev}. Surface scaling in sandpiles
can be different from the bulk one and can also depend on the type of boundary
conditions (b.c.) considered. In those models the obvious alternative
to dissipative b.c. are conditions in which part of the border does not
dissipate grains\cite{STC}.
For evolution models, which do not dissipate particles,
it is not known whether boundary conditions could influence scaling at
the border and, in case, what should correspond to conservation.
These are issues we address in the present article.

Especially in the perspective of obtaining exact results,
the study of boundary scaling should
represent an important step towards a deeper and more complete
theoretical understanding of SOC.
In the present article we show that the $M=\infty$ model
of Ref.\cite{BP}, if
considered in the presence of boundaries, provides a microscopic realization
of the inhomogeneous BP introduced in \cite{CTS}. Besides
generalizing immediately to this model results known for BP,
this opens new possibilities
of both analytical and numerical investigation. In particular, by extending
methods used previously for the bulk\cite{BP}, we are able to
compute exactly the
asymptotic two--point correlator when it involves a point on the boundary
in the $M=\infty$ limit. These results, and an accurate numerical analysis
of time correlations at the boundary, allow to draw a complete scenario
of the scalings obeyed by boundary and bulk exponents of the system.
Boundary avalanches have different scaling properties for
different boundary conditions. However, even in the case
of dissipative border, in which
exponents differ from those in the bulk,
space and time fractal dimensions of the avalanches
remain unaltered and satisfy the same basic scaling relations.

This paper is organized as follows. In the next section, after introducing
the semi--infinite $M$--component model, we discuss its relation with
inhomogeneous BP in the $M \to \infty$ limit and derive a number of
analytical and numerical results for size and spatial distribution
properties. The third section is devoted to a discussion 
of exponents related to the time recurrence of
activity at the
border. The last section contains general conclusions.
\section{Semi--infinite BS evolution chain with $M=\infty$ components.}

The possible relevance of the BS model for evolution, as revealed, e.g., by
paleonthological records\cite{GE}, has been already discussed
in the literature. Here we regard this model and its variants as an
interesting mathematical framework within which SOC dynamics can be studied.

We consider an open chain of species, labelled by an index $i=1,2,3..$.
Each species is characterized by $M$ independent parameters (traits)
${x_i}^{\alpha}$, ($\alpha=1,2,..,M$, $0<{x_i}^{\alpha}<1$), which somehow
quantify the ability of the species to survive in connection with $M$
different tasks it has to perform in the eco--system. The closer
${x_i}^{\alpha}$ to $1$, the more the ability connected to
the $\alpha$--th task, and thus the
higher the chance that the species avoids mutation.

The dynamics goes as follows. At every time step the smallest ${x_i}^{\alpha}$,
i.e. the weakest among the traits of all species, is identified and
replaced by $1$.
Each one of the species
which are neighbors along the chain of that, $i_{min}$, with minimum
$x^{\alpha}$, get replaced one of their traits (chosen at
random among the $M$ possible ones) with new random numbers extracted
independently and with
uniform probability in the interval $(0,1)$. A new minimum is then searched
and this proceeds so that on long times the system
self--organizes itself in a stationary state with all
${x_i}^{\alpha}$ uniformly distributed in an interval $(\lambda_c, 1)$.

A $\lambda$--avalanche is identified with a sequence
of mutations starting at site $i_{min}$
with $x_{i_{min}}^{\alpha}=\lambda$, and continuing
until the current minimum $x^{\alpha}$ remains below $\lambda$. We call $s$
(duration of the avalanche) the total number of minima with value below
$\lambda$ obtained during the avalanche.

Rather then considering a translation invariant situation as in Ref.\cite{BP},
we think here to a semi--infinite chain, with suitable b.c..
The probability that in the stationary state a
$\lambda$--avalanche has size $s$, will thus depend on the site $j$,
$(j=1,2,...)$,
where the avalanche started. Omitting to explicitate the
$\lambda$--dependence, we indicate such probability by
$ P_j(s)$. The above dynamical rules, for $M \rightarrow \infty$,
lead to write
\begin{eqnarray}
P_j(s+1) &=& \lambda(1-\lambda)[P_{j+1}(s)
+P_{j-1}(s)] +  \nonumber \\
& & \lambda^2 \sum_{s^\prime = 0}^s P_{j+1}(s^\prime )
P_{j-1}
(s-s^\prime )
\,\,\,\,\,\,\,\,\,\,  j > 1
\label{prob}
\end{eqnarray}
Eq.(\ref{prob}) is derived on the basis of the same
considerations made in Ref.\cite{BP}.
In first place Eq.(\ref{prob}) reflects the fact that, with our
dynamical rules,
the starting active site (site $j$ for which $x_j^{\alpha}=\lambda=$
absolute minimum) can activate one of its neighbors with probability
$\lambda(1-\lambda)$, or both with probability $\lambda^2$.
In the two cases, of course, one or two indipendent avalanches follow,
respectively, and the global avalanche grows up to a total of $s+1$ activated
sites.
The above independence, which allows
to write Eq.(\ref{prob}) in such form, holds in the $M \rightarrow
\infty$ limit,
which is implicitly assumed here. Indeed, only with
$M = \infty$ the evolution of an avalanche is completely
unaffected by the fact that a given site has been previously
involved in the same, or another, avalanche. The effects of this kind of
conditions can indeed be seen to amount to corrections of the order
${1 \over M}$, or higher, in the equations of motion.
Eq.(\ref{prob}) needs of course to be complemented with b.c..
These conditions
can be written in different ways. A first possibility is:
\begin{equation}
P_1(s+1)=\lambda P_2(s)
\label{bound1}
\end{equation}
This means that, once the boundary site $1$ becomes active, it can then
activate only site $2$ (site 0 does not exist), and this occurs with the usual
modalities. An alternative boundary condition is:
\begin{eqnarray}
P_1(s+1)&=&\lambda (1-\lambda )[P_1 (s)
+P_2(s)] + \nonumber \\
& & \lambda^2 \sum_{s^\prime =0}^s P_1(s^\prime )P_2
(s - s^\prime )
\label{bound2}
\end{eqnarray}
Eq.(\ref{bound2}) means that for the boundary site, when active, the rule of
setting
$x^\alpha=1$, valid for $j>1$, does not apply. On the contrary,
$j=1$ and $j=2$ get now the random replacement of one of their
traits, as sites $j-1$, and $j+1$ in the bulk (see Eq.(\ref{prob})). In other
words, the role of the missing site $j=0$ is now played by the site $j=1$
itself.

It is straightforward to recognize that, up to minor modifications due
to the convention assumed here of replacing by $1$ the $x$ associated to
the minimum trait, Eqs.(\ref{prob}) and (\ref{bound1}) have the same
structure as those describing the inhomogeneous BP in
$1D$ of Ref.\cite{CTS}. By introducing generating functions
${\tilde{P}}_i(x)=\sum_{s=0}^{\infty}P_i(s)x^s$, $i=1,2,..$,
it was found there that ${\tilde{P}}_1(x)\sim
1+c(1-x)^{1-7/4}$, for $x \to 1^{-}$, when $\lambda=1/2$.

\begin{figure}
\centerline{ \psfig{file=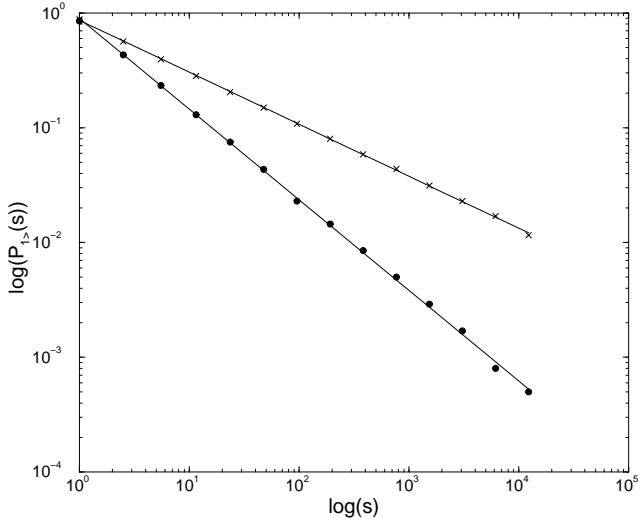,height=7cm,angle=-90} }
\caption{The log-log plots of the integrated $s$ distribution of avalanches starting at
site 1,  $P_{1>}(s)= \int_s^{\infty} P(x) dx$ with b.c. 
(2) (dots) and (3) (crosses).}
\label{FIG01}
\end{figure}

This value
of $\lambda$ is such to imply an average number $ 2 \lambda(1-\lambda)+
2 \lambda^2=1$ of sites activated by each active site in the bulk and
coincides with
$\lambda_c$\cite{BP}. The
average number of sites activated by the border site $1$
is instead less than unity, according to Eq.(\ref{bound1}). All this means
that, for $\lambda=\lambda_c=1/2$, ${P}_1 \sim s^{-\tau'}$,
with $\tau'=7/4$, when $s\to \infty$. This has to be compared
with the result $P_{\infty} \sim s^{-3/2}$\cite{BP} holding
when the site where the avalanche starts is chosen in the bulk, and
implying a mean field bulk exponent $\tau=3/2$\cite{MF}.

We indicate by $N(j,r)$ the probability that, at $\lambda=\lambda_c=1/2$, an
avalanche started at site $j$ never reaches site $r \ge j$. We are
interested in the behavior of $N$ for $j=1$ and large $r$, which is in turn
related to the asymptotic radial distribution of avalanches starting
at the border of the chain. The Markovian nature of avalanche evolution
allows to write
\begin{eqnarray}
N(j,r)&=&\frac{1}{4}+\frac{1}{4}[N(j+1,r)+N(j-1,r)]+ \nonumber \\
& & \frac{1}{4}N(j+1,r)N(j-1,r),
\label{n1}
\end{eqnarray}
for $2<j<r-1$. The b.c. in Eq.(\ref{bound1}) imply
\begin{equation}
N(1,r)=\frac{1}{2}+\frac{1}{2}N(2,r),
\label{n2}
\end{equation}
while, since obviously $N(r,r)=0$,
\begin{equation}
N(r-1,r)=\frac{1}{4}+\frac{1}{4}N(r-2,r).
\label{n3}
\end{equation}
Let us put  now $N(j,r)=1-f(j,r)$, so that $f$ represents the
probability that an avalanche starting in site $j$ reaches site $r$.
From the above equations follows:

\begin{figure}
\centerline{ \psfig{file=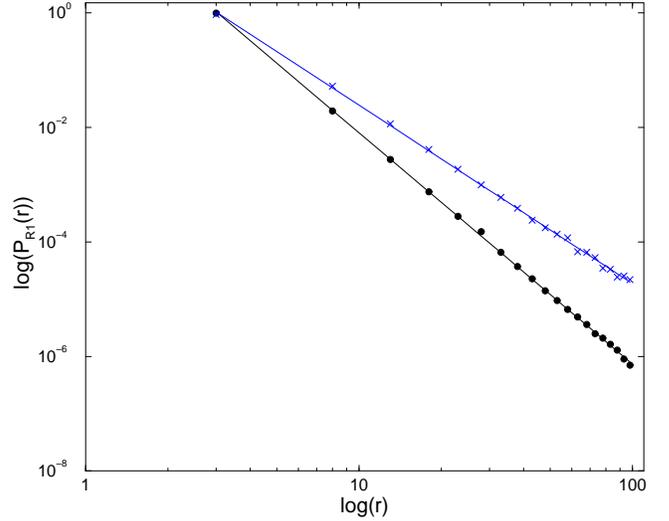,height=7cm,angle=-90} }
\caption{Log-log plots of $P_{R1}(r)$ for b.c. (2) (dots) and b.c. (3) (crosses).}
\label{FIG02}
\end{figure}

\begin{eqnarray}
\Delta f(j,r)&=&\frac{1}{2} f(j-1,r) f(j+1,r)
\,\,\,  1<j<r-1 \nonumber\\
\Delta f(r-1,r)&=&\frac{1}{2}-\frac{3}{4}f(r-2,r) \nonumber\\
\Delta f(1,r)&=&\frac{1}{2}f(2,r),
\label{n4}
\end{eqnarray}
where $\Delta f(k,r)$ is the discrete Laplacian of $f$ at site $k$.
Since we are interested in the large $r$ behavior, we can pass to a
continuum limit, introducing the variable $z=(j-1)/r$. By putting
$f(j,r)=y(z)$, we obtain from Eq.(\ref{n4})
\begin{equation}
\frac{y''(z)}{r}=\frac{y(z)}{2};
\,\,\,\, 0<z<1,
\label{n5}
\end{equation}
with boundary conditions $y'(1)/r=-3y(1)/4+1/2$ and $y'(0)/r=y(0)/2$.
Eq. (\ref{n5}) can be integrated and, after some algebra, one finds:
\begin{eqnarray}
f(1,r) =y(0) & \simeq &
\frac{6}{r^3}\left[\int_0^{\infty}\frac{dx}{\sqrt{x^3+1}}
\right]^3  \nonumber\\
& \simeq & \frac{2}{9}
\frac{1}{r^3}\left[\frac{\Gamma(1/3)\Gamma(1/6)}
{\Gamma(1/2)}\right]^3 \simeq \frac{20.14..}{r^3}.
\label{n6}
\end{eqnarray}
If the probability that a critical avalanche starting at site $1$
reaches site $r$ is $P_{R1}(r)\propto r^{-\tau_R'}$, we conclude
$\tau_{R}'=4$, from the fact that $f(1,r)\propto \int_{r}^{\infty}
P_{R1}(x) dx$.
The above derivation extends the approach of Ref.\cite{BP}, which
yielded $\tau_{R}=3$ for the bulk $P_{R}(r)$.

The presence of a border like that specified
by the b.c. in Eq.(2), should be expected to make the avalanche propagation
more difficult compared to the bulk situation. Thus, the result
$\tau_R'=4$ is physically sound compared to $\tau_R=3$.
The fact that
the radial probability distribution function
decays with a different exponent when the starting point is at the
border with b.c. (\ref{bound1}), is qualitatively consistent with what we
know of the two--point
correlator at an equilibrium critical point when one of the points
is fixed at the boundary and not in the bulk\cite{surface}.

The exact results above for $\tau'$ and $\tau_{R}'$ allow to draw a
first conclusion on the space fractal dimension, $D'$, of avalanches
starting at a border with b.c. (2). Assuming
$s \propto r^{D'}$ for such an
avalanche, leads to $D'(1-\tau')=1-\tau_{R}'$, which follows from
$P_1(s) ds = P_{R1}(r) dr$. Thus, $\tau'=7/4$ and $\tau_{R}'=4$ imply
$D'=4$. This dimension coincides with the bulk one, $D$, which satisfies
the same kind of relation $D(1-\tau)=1-\tau_R$\cite{BP}.
So, at the boundary,
there is no distinct space fractal dimension for these
avalanches, in spite of the different $\tau$-exponent. 

We verified numerically the above result for $\tau'$, by simulating the
model on open finite chains of length $N=200$, with
$M=100$ components.
Fig. 1 reports our finite size data  referring to avalanches
starting
at the border with b.c. given by Eq.(2). The distribution is well
compatible with the expected $\tau'=7/4$ ( we estimated $\tau'=1.78 \pm 0.04$).
Direct simulation allows us also to investigate the implications
of b.c. (3), which, as far as $\tau'$ is concerned,
escape analytical control. For these b.c., which keep equal to $1$
the avarage number of sites activated by the border one, we find
numerically $\tau'=1.46\pm 0.04$, compatible with $\tau'=\tau=3/2$ (Fig.1).
Since the result  $\tau'=7/4$ should hold for the inhomogeneous BP
as long as $2\lambda(1-\lambda)+\lambda^2<1$\cite{CTS}, we conclude that
$\tau'=\tau=3/2$ is peculiar of b.c. (3). Consistently one can also show
that, with b.c. (3), $\tau_R'=\tau_R=3$ exactly. Thus, also $\tau_R'$ is
restored to its bulk value by b.c. (3).
In the SOC context,
similar results were previously conjectured, on numerical basis, for
the Abelian sandpile in two dimensions\cite{STC}. Indeed, for
that model
border avalanches appeared to possess a toppling distribution exponent
rather close to the bulk value for conservative border,
while with boundary dissipation a different $\tau'$ applied\cite{STC}.
Border dissipativity in a BS evolution
model should then be associated to the fact that the average number
of sites activated by the extremal site is less than the critical bulk
value.

Fig. 2 illustrates a numerical determination of $\tau'_R$ for avalanches
starting with
dissipative b.c. (2). We obtain $\tau'_R=4.05 \pm 0.05$ well consistent
with our exact result. With b.c. (3) we get $\tau_R'=3.02\pm 0.08$, compatible
in this case with the exact $\tau_R'=\tau_R=3$ (Fig.2).

\section{Time fractal properties.}

Avalanches of a BS model posses also time fractal properties,
revealed, e.g., by the distribution of first return times of the
activity in a given site (time being measured by the number of
minima which get replaced during the avalanche). Some general
relations among exponents connected with time and space
fractal behavior in the bulk\cite{PMB} can be easily derived by
arguing as follows.

\begin{figure}
\centerline{ \psfig{file=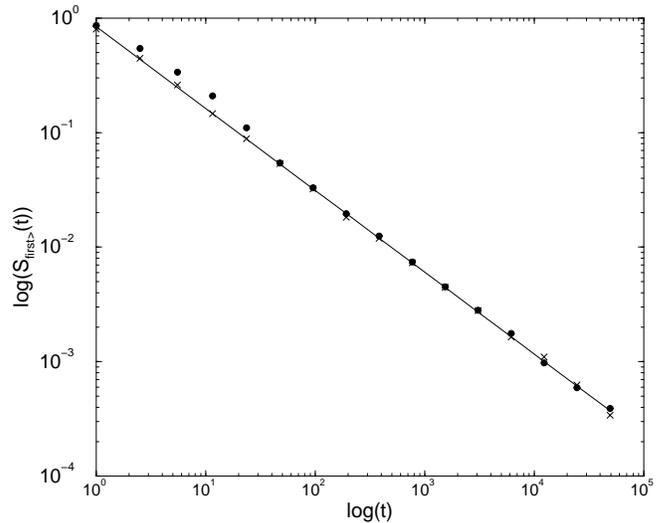,height=7cm,angle=-90} }
\caption{Log-log plots of the integrated first return time disrtibutions in the
case of b.c. (2) (dots) and b.c. (3) (crosses). }
\label{FIG03}
\end{figure}

If we define as $S_{first}(t)$ the probability distribution of
first return times in a given site, and call $n(T)$ the total number
of returns in a lapse of time $T$, we expect $n(T)\propto T^{\tilde
{d}}$, where $\tilde{d}$ is a time fractal dimension, and
\begin{equation}
\frac{T}{n(T)}\propto \int_1^{T} S_{first}(t) t dt
\label{n7}
\end{equation}
Upon putting $S_{first}(t)\propto t^{-\tau_{first}}$, we get
$\tilde{d} =\tau_{first}-1$. Let us then call $S_{all}(t)\propto
t^{-\tau_{all}}$ the distribution
of times for all subsequent returns in a given site. We clearly have
$\int_1^{T} S_{all}(t)dt\propto T^{\tilde{d}}$,
so that $\tilde{d}=1-\tau_{all}$
and $\tau_{first}+\tau_{all}=2$. At this point, to link space and
time fractal properties is sufficient to consider an avalanche as
made of a total of $s$ activated sites within a $d$--dimensional hyperspherical
region of radius $r$ such that $s\propto r^D$. If the avalanche has
time duration $t$, we must have $s\propto r^d n(t)=r^d n(r^z)$,
where $z$ is an exponent connecting space and time ($t\propto r^z$).
The last relation treats all lattice sites within the sphere as
equivalent, as far as the return of activity is concerned. Now, since in
our model $s=t$ by definition, $z=D$ also applies. Eventually, one finds:
\begin{equation}
\tau_{first}= 2-\frac{d}{D}
\label{n8}
\end{equation}
This relation was first proposed in Ref.
\cite{PMB} for bulk exponents. The present derivation at first
sight looks applicable also to avalanches starting at the border.
The $d=1$, $M=\infty$ BS model is an ideal
context in which to test  its validity. In Ref.\cite{BP} a numerical
estimate of $\tau_{first}$ was obtained which turned out to be
well compatible with the
value implied by relation (11) ($\tau_{first}=7/4$). We made a similar
determination of $\tau_{first}'$ for avalanches starting at both
dissipative and nondissipative
borders. The data are plotted in Fig.3 , where one can
clearly appreciate that the same values of $\tau_{first}'$ apply
in the two cases.  Indeed, for conservative b.c. (Eq. (3)) we estimate
$\tau_{first}'=1.71 \pm 0.03\simeq 7/4$.
With dissipative b.c. there appears to be
a longer transient before the asymptotic time scaling behavior
establishes. However,
we estimated $\tau_{first}'= 1.71 \pm 0.04$, clearly compatible again with
$7/4$. In both cases, of course, the bulk avalanches have a distribution
with $\tau_{first}\simeq 7/4$.

Our results indicate that, like the space fractal dimension $D'$,
also the time fractal dimension $\tilde{d}'$ of boundary avalanches is the same
as its bulk counterpart, with all b.c., and Eq. (11) is always satisfied.

In Ref. \cite{CTS} a simulation of the $d=1$, $M=1$ BS model yielded a
$\tau'_{first}$ sensibly different from $\tau_{first}$. If such a
discrepancy would be confirmed by more systematic and asymptotic
determinations, one should suspect that the identity of $D'$ and $D$,
or even the validity of some scaling relations like Eq.(11),
are somehow granted here by the peculiar, classical character of the
$M=\infty$ model. Anyhow, in such a case, a more complex scaling scenario
would certainly apply to the model of Ref. \cite{CTS}.

\section{ Conclusions.}

The $M$--component BS model in the limit $M\to \infty$ is an
interesting theoretical laboratory for testing properties of the
SOC state. With the present work we were able to compute
analytically in $d=1$ the exponents $\tau'$ and $\tau_R'$ referring,
respectively, to size distribution and space correlation of avalanches
starting at a border specified by b.c. (2). This extends previous results in
Ref.\cite{BP}, which referred exclusively to bulk properties. In addition
our formulation allowed to establish a direct link between this model
and the inhomogeneous branching process discussed in Ref.\cite{CTS}.

The results $\tau'=7/4$ and $\tau_R'=4$ show that the space fractal dimension
$D'$ of border avalanches with b.c. (2) remains equal to the bulk one
($D'=D=4$), in spite of the change of these exponents.
Complemented with numerical tests, these results showed the
existence of a
clear--cut distinction between the b.c. in Eq. (2) and those expressed by
Eq. (3). In analogy with the physics of sandpile models, we were led to
call b.c. (2) dissipative, due to the fact that, in force of them, the
boundary site, on average, is able to transmit activity to less than one
site, even if the bulk is critical. This dissipativity is responsible for
boundary values of the exponents $\tau'$ and $\tau_R'$
different from the bulk ones. On the other hand,
when b.c. are conservative in the sense speciefied by Eq. (3), the existence
of a geometrical boundary does not seem
enough to determine a scaling different from the bulk one.

To border dissipation, which for models like sandpiles is a necessary
condition for the very establishment of the stationary SOC state,
could be given here a precise meaning also in the context of
a model with extremal dynamics. In this model dissipation reveals
an essential ingredient for the existence of
peculiar boundary scaling, distinct from the bulk one. Indications that
this could be a general feature of the SOC state come also from previous
numerical results for sandpiles \cite{STC}.

Our study of the return of activity at the border site revealed that
dissipativity does not determine a new boundary $\tau_{first}'$ exponent,
consistent with Eq. (11) and with the fact that, like $D'$,
for boundary avalanches also $\tilde{d}'$
remains unaltered with respect to its bulk value.

This contrasts, if confirmed by further analysis, awaits to be elucidated.

ALS wishes to thank MIT, and M. Kardar in particular, for hospitality
within the INFN(Italy)-MIT ``Bruno Rossi'' exchange program. Partial
support from the European Network Contract N. ERBFMRXCT980/83 is also
acknowledged.

\end{multicols} \widetext

\begin{references}
\bibitem{BTW} P. Bak, C. Thang, and K. Wiesenfeld, Phys. Rev. Lett.
{\bf 59}, 381 (1987); Phys. Rev. A {\bf 38},
364 (1988); D. Dhar, Phys. Rev. Lett. {\bf 64}, 1613 (1990).
\bibitem{Olami} Z. Olami, H. J. S. Feder, and K. Christensen,
Phys. Rev. Lett. {\bf 68}, 1244 (1992); K. Ito, Phys. Rev. E {\bf 52}, 3232
(1995).
\bibitem{Sneppen} K. Sneppen, Phys. Rev. Lett. {\bf 69}, 3539 (1992).
\bibitem{BS} P. Bak and K. Sneppen, Phys. Rev. Lett. {\bf 71},
4083 (1993);
\bibitem{Maslov} See, however, S. Maslov, Phys. Rev. Lett. {\bf 77},
1182 (1996); M. Marsili, P. De Los Rios, and S. Maslov, Phys. Rev.
Lett. {\bf 80}, 1457 (1998).
\bibitem{PMB} M. Paczuski, S. Maslov, and P. Bak, Phys. Rev. E
{\bf 53}, 414 (1996).
\bibitem{MF} H. Flyvbjerg, K. Sneppen and P. Bak, Phys. Rev. Lett.
{\bf 71}, 4087 (1993); J. de Boer, B. Derrida, H. Flyvbjerg, A. D. Jackson,
and T. Wettig, Phys. Rev. Lett. {\bf 73}, 906 (1994).
\bibitem{CTS} G. Caldarelli, C. Tebaldi, and A. L. Stella,
Phys. Rev. Lett. {\bf 76}, 4983 (1996).
\bibitem{Lise} At a purely numerical level, one possible candidate
has been recently identified in a suitable random neighbor model
of earthquakes belonging to the family of sandpiles\cite{BTW}). See
S. Lise and A. L. Stella, Phys. Rev. E {\bf 57}, 3633 (1998).
\bibitem{BP} S. Boettcher and M. Paczuski, Phys. Rev. Lett. {\bf 76},
348 (1996); Phys. Rev. E{\bf 54}, 1082 (1996).
\bibitem{STC} A. L. Stella, C. Tebaldi, and G. Cardarelli,
Phys. Rev. E {\bf 52}, 72 (1995).
\bibitem{Priezzhev} E. V. Ivashkevich, D. K. Ktitarev and V. B. Priezzhev,
J. Phys. A {\bf 27}, L585 (1994).
\bibitem{GE} S. J. Gould and N. Eldredge, Nature (London) {\bf 366},
223 (1993).
\bibitem{surface} K. Binder in {\it Phase Transitions and Critical Phenomena},
Vol. {\bf 8 } p. 1, Ed. by C. Domb and J. L. Lebowitz (Academic, London, 1983).
\end{references}
\end{document}